\documentclass[12pt]{article}
\usepackage{graphicx}
\usepackage{setspace}
\begin{document}
\bibliographystyle{unsrt}
\def\ra{\rangle}
\def\la{\langle}
\def\aao{\hat{a}}
\def\aaot{\hat{a}^2}
\def\aco{\hat{a}^\dagger}
\def\acot{\hat{a}^{\dagger 2}}
\def\ano{\aco\aao}
\def\bao{\hat{b}}
\def\baot{\hat{b}^2}
\def\bco{\hat{b}^\dagger}
\def\bcot{\hat{b}^{\dagger 2}}
\def\bno{\bco\bao}
\def\beqn{\begin{equation}}
\def\eeqn{\end{equation}}
\def\bear{\begin{eqnarray}}
\def\eear{\end{eqnarray}}
\def\cdott{\cdot\cdot\cdot}
\def\bcen{\begin{center}}
\def\ecen{\end{center}}
\def\nbar{\bar{n}}
\def\eps{\epsilon}
\def\hrho{\hat{\rho}}
\def\rhom{\hat{\rho}_m}
\def\rhot{\hat{\rho}_t}
\def\rhod{\hat{\rho}_d}
\bcen
{\Large Superposed Coherent States improve fidelity of NOON states  generated in post-selection}\\
\vspace{1in}
S. Sivakumar\\Materials Physics Division\\ 
Indira Gandhi Centre for Atomic Research\\ Kalpakkam 603 102 INDIA\\
Email: siva@igcar.gov.in
\ecen
\begin{abstract}
 Mixing the squeezed vacuum and coherent state in a beam-splitter generates NOON states  of the electromagnetic field, but with fidelity less than unity.  In this note it  is shown that the output of the beam-splitter generates the NOON states with unit fidelity if superposed coherent states are used instead of the squeezed vacuum.  Also, the post-selection probability of NOON states is shown to be higher.     
\end{abstract}
\newpage
\section{Introduction}
  
Interferometers  are useful in measuring  the phase changes introduced by a non-opaque object.  Usually, the object that changes the phase of the wave is introduced in one of the arms of the interferometer.  The accuracy in the estimation of the relative phase $\delta\phi$  is limited by the standard quantum limit (SQL)  $\delta\phi=1/\sqrt{N}$, where  $N$ is the number of particles used in the measurement.  
The accuracy is improved by increasing the number of particles.  However, quantum features make it possible to achieve higher accuracy than the SQL, for a given number of particles\cite{ou}.  For instance, on using the nonclassical,  squeezed light\cite{gerryknight} in the input ports of Mach-Zehnder interferometer (MZI), the sensitivity approaches $1/N$\cite{caves}.   It has been realized that nonclassical correlations and entanglement are of relevance in surpassing the SQL\cite{ou, holland, benatti}.  In that context, an important class of bipartite entangled states that is useful is the NOON state\cite{sanders, boto, lee, dowlingcp}, 
\beqn
\vert\hbox{NOON}\ra=\frac{1}{\sqrt{2}}\left[\vert N,0\ra+\vert 0,N\ra\right].
\eeqn
Here $\vert N,0\ra$ and $\vert 0,N\ra$ correspond to the product states $\vert N\ra\vert 0\ra$ and $\vert 0\ra\vert N\ra$ respectively.  It is possible to surpass the SQL in phase measurements if NOON states are used.  Apart from their significance in phase measurements, the NOON states have been studied in the context of spatial super-resolution surpassing the Rayleigh limit\cite{boto}, multiphoton interferometry\cite{ourev} gravity wave detection\cite{gravitynoon} and quantum metrology in general\cite{slloyd,pracqm}.  
Because of their relevance in very many contexts, generation of optical NOON states has acquired a great significance.  Many proposals have been put forth and some of them have been realized in the laboratories\cite{zeilinger, mitchell, nagata, kolkiran}.   
 Recently, these states have been generated by the method of post-selection \cite{ono} on the output of a beam-splitter with properly chosen input states\cite{afek}.  The output state is not exactly the NOON state, but the N-photon component of the output state is almost like a NOON state.  This is a significant feature of this experiment.   The quality of the  output is expressed  in terms of fidelity $F$, which is defined as the overlap between the NOON state and the normalized N-photon component of the output state from the beam-splitter. If $\sum_{n,m=0}^\infty C_{nm}\vert n,m\ra$ is the bipartite state at the output of the beam-splitter, then 
\beqn
F=\vert\la\hbox{NOON}\vert\sum_{n=0}^N C_{n,N-n}\vert n,N-n\ra\vert^2=\frac{\vert C_{0,N}+C_{N,0}\vert^2}{2}.
\eeqn 

Parameters such as the amplitudes of the input  states are optimized for each $N$ to achieve maximum fidelity.

In the experiment of  Afek {\it et al}\cite{afek}, a beam-splitter is used to generate NOON state.  The input for the beam splitter is the product state of the  squeezed vacuum (SV) $\vert r\ra$ and the coherent state (CS) $\vert\alpha\ra$ of amplitudes $r$ and $\alpha$ respectively.  The  number state expansions of these states are, 
\beqn
\vert r\ra=\frac{1}{\cosh r}\sum_{n=0}^\infty(-\tanh r)^n\frac{\sqrt{(2n)!}}{2^nn!}\vert 2n\ra.
\eeqn
and  
\beqn
\vert\alpha\ra=\exp(-\frac{\vert\alpha\vert^2}{2})\sum_{n=0}^\infty\frac{\alpha^n}{\sqrt{n!}}\vert n\ra.
\eeqn
Here, the amplitude $r$ is real and $\alpha$ is complex.   With $\vert r\ra\vert\alpha\ra$ as the input to the beam-splitter, the output contains all the NOON states as its components, in addition to non-NOON state components.  The two input modes of the beam-splitter are denoted as $a$-mode and $b$-mode respectively.      

\section{Ideal input state to generate NOON states}

In this section, the ideal input states for  the experimental arrangement of \cite{afek} to generate NOON state by post-selection is discussed.  The arrangement generates entanglement by  a beam-splitter.  The quantum state of the bipartite output from a lossless beam splitter  is given by the action of the unitary operator,
\beqn
U_{bs}(\gamma)=\exp(\gamma\aao\bco-\gamma^*\aco\bao),
\eeqn
on the input bipartite state\cite{gerryknight}.  Here  $\gamma=i\pi/4$ for lossless, 50-50 beam splitter\cite{gerryknight}.  The operators $\aco$ and $\aao$ are the creation- and annihilation-operators for the $a$-mode.  The corresponding operators for the $b$-mode are $\bco$ and $\bao$ respectively.    The suitable  input states for the beam-splitter to generate NOON states at the output are  obtained by observing that 
\beqn
\vert\hbox{NOON}\ra=U_{bs}(\gamma)U^\dagger_{bs}(\gamma)\vert\hbox{NOON}\ra,
\eeqn
where the adjoint operator $U^\dagger_{bs}(\gamma)=U_{bs}(-\gamma)$.  
This relation  implies that the apt input to produce NOON states is $U_{bs}(-\gamma)\vert\hbox{NOON}\ra$.   To analyze further, the bipartite number state expansion of the input state is useful.  To this end, it may be noted that the operators $\aco\bao$, $\aao\bco$ and $\ano-\bno$ form a closed algebra under commutation. 
Defining $K_+=\aco\bao, K_-=\aao\bco$ and $K_0=(\ano-\bno)/2$, we have
\beqn
[K_\pm,K_0]=\pm K_\pm, ~~~~~[K_-,K_+]=K_0.
\eeqn
These commutation relations define the SU(2) algebra\cite{eberlysu2, buzeksu2}.  Consequently, the operator $U^\dagger_{bs}(\gamma)$ has the following disentangled forms\cite{disent}
\bear
U^\dagger_{bs}(\gamma)&=&\exp(p\aco\bao)\exp[q(\ano-\bno)]\exp(r\aao\bco)\\
&=&\exp(p\aao\bco)\exp[q(\bno-\ano)]\exp(r\aco\bao),
\eear
Setting $\gamma=\vert\gamma\vert\exp(-i\theta)$, the parameters $p$, $q$ and $r$ are  $\exp(-i\theta)\tan\vert\gamma\vert,~2\log\sec\vert\gamma\vert$ and $-\exp(-i\theta)\tan\vert\gamma\vert$ respectively.  
 Employing the disentangled form of  the  unitary operator $U^\dagger_{bs}(\gamma)$, the number state expansion of the  required input state is obtained to be 
\beqn\label{ginp}
U^\dagger_{bs}\vert\hbox{NOON}\ra=\frac{\exp(-qN/2)}{\sqrt{2}}\sum_{k=0}^N\sqrt{N\choose k}(p^k+p^{N-k})\vert N-k,k\ra.
\eeqn
This is the input state that a beam-splitter converts to a NOON state.   It is of interest to note that the required input state can be viewed as a superposition of two SU(2) coherent states: one generated from the state $\vert N,0\ra$ and the other generated from the state $\vert 0,N\ra$.

	For a 50-50, lossless, symmetric  beam-splitter, $\gamma=i\pi/4$  and  the values of  $p$, $q$ and $r$ are $-i,~\log 2$ and $-i$ respectively.   With this choice of $\gamma$, 

\beqn\label{inp+}
U^\dagger_{bs}(i\frac{\pi}{4})\vert\hbox{NOON}\ra=\frac{1}{\sqrt{2}^{N-1}}\sum_{k=0}^{N/2}(-1)^k\sqrt{N\choose{2k}}\vert N-2k,2k\ra,
\eeqn
if $N$ is an even  multiple of two, for example, $N=4,8,12,\cdots$.  If $N$ is an odd multiple of two [$N=2(2m+1)$], then the required input state is
\beqn\label{inp-}
U^\dagger_{bs}(i\frac{\pi}{4})\vert\hbox{NOON}\ra=\frac{1}{\sqrt{2}^{N-1}}\sum_{k=0}^{N/2-1}(-1)^k\sqrt{N\choose{2k+1}}\vert N-2k-1,2k+1\ra.
\eeqn 
It is to be pointed out that if $N$ is an odd multiple of two, then the required input involves bipartite number states with respective odd numbers of photons.  Such a combination is not possible if the SV is used as one of the input fields since the state is a linear superposition of even number states only.  
The finite superposition given in Eqns. \ref{inp+} and \ref{inp-} are the respective  input  states to generate the NOON states corresponding to even and odd multiples of two.   

 But these finite superpositions are difficult to generate experimentally for large $N$. Alternately,  it is possible to use bipartite states of the electromagnetic field whose $N$-photon component is similar to the form of the ideal input states.  This is the idea suggested in \cite{ono} and subsequently realized\cite{afek}.   For proper choice of the amplitudes of the input states, the maximum achievable fidelity is 0.94.  This has been experimentally  implemented to generate NOON states upto $N=5$. 

\section{Superposed coherent states versus squeezed vacuum}

	It is interesting to see if the fidelity can be increased to unity by a suitable choice of input states.  This, in turn, means that the bipartite input to the beam-splitter should have N-photon component in the form given in Eq. \ref{inp+} and Eq. \ref{inp-} depending on whether $N$ is an even or odd multiple of two\cite{urias}.  Like the SV, even coherent states (ECS) are constructed as a superposition of the even number states \cite{yurkestoler,schleich,eocs} of the electromagnetic field.  The ECS, denoted by $\vert\beta,+\ra$, of complex amplitude $\beta$ are obtained as the symmetric combination of the coherent states $\vert\beta\ra$ and $\vert -\beta\ra$,
\beqn
\vert\beta,+\ra=
\frac{1}{\sqrt{\cosh\vert\beta\vert^2}}\sum_{n=0}^\infty\frac{\beta^{2n}}{\sqrt{(2n)!}}\vert 2n\ra.
\eeqn
Similarly, odd coherent states (OCS), denoted by $\vert\beta,-\ra$,  are defined as the antisymmetric combination of the coherent states of amplitudes $\beta$ and $-\beta$,
\beqn
\vert\beta,-\ra=
\frac{1}{\sqrt{\sinh\vert\beta\vert^2}}\sum_{n=0}^\infty\frac{\beta^{2n+1}}{\sqrt{(2n+1)!}}\vert 2n+1\ra.
\eeqn
Both the ECS and OCS have been studied in the context of the optical  Schrodinger cat states. In addition, these states posses many other nonclassical features such as squeezing, sub-Poissonian statistics, etc.  Of importance in the present context is that such superpositions of the propagating light fields can be generated  in experiments \cite{schcat,ocs,schkitten,glancy}.

Now, the suitability of the ECS  and OCS in the NOON state generation is discussed.     The $N$-photon component of the  product state $\vert\beta,+\ra\vert\alpha\ra$  is
\beqn\label{ecs2}
\vert\beta +,\alpha\ra_{N,e}\propto\sum_{n=0}^{N/2}\sqrt{N\choose{2k}}\left[\frac{\alpha}{\beta}\right]^{2k}\vert N-2k, 2k\ra,
\eeqn
if $N$ is even, while it is
\beqn
\vert\beta +,\alpha\ra_{N,o}\propto\sum_{k=0}^{N/2-1}\sqrt{N\choose{2k+1}}\left[\frac{\alpha}{\beta}\right]^{2k+1}\vert N-2k-1, 2k+1\ra,
\eeqn
if $N$ is odd.  If the input is the OCS-CS $\vert\beta,-\ra\vert\alpha\ra$, the N-photon components corresponding to even N is
\beqn
\vert\beta -,\alpha\ra_{N,e}\propto\sum_{k=0}^{N/2}\sqrt{N\choose{2k+1}}\left[\frac{\alpha}{\beta}\right]^{2k+1}\vert N-2k-1, 2k+1\ra,
\eeqn
while the odd N component is 
\beqn\label{ocs3}
\vert\beta -,\alpha\ra_{N,o}\propto\sum_{n=0}^{(N-1)/2}\sqrt{N\choose{2k}}\left[\frac{\alpha}{\beta}\right]^{2k}\vert N-2k, 2k\ra.
\eeqn
Comparison of Eqs. \ref{ecs2} and \ref{ocs3}  with  Eqs. \ref{inp+} and \ref{inp-} respectively makes it clear that the N-photon component of the input states are precisely of the required form to generate the NOON states if $\alpha=i\beta$. This, in turn, means that the NOON state fidelity of the output state is unity.    Thus, ideal NOON states of even $N$ can be obtained by mixing coherent states with Schrodinger cat states.  The discussion so far has been restricted to generating those NOON states where $N$ is even.  It may be noted that with the choice of $\gamma$ used here,  the fidelity of NOON state is always less than unity if $N$ is odd.

   Though the fidelity defined earlier gives a measure of the "NOON"ness of the N-photon component of the output state, the probability of NOON state occurrence is given by the overlap between the input state to the beam-splitter and the ideal input state.  This quantity can also be viewed as the overlap between the NOON state and the realized output state.   As indicated before, the ideal input state to NOON states with $N$ as an even multiple of two is contained in the state $\vert\alpha +\ra\vert i\alpha\ra$.   The corresponding overlap is 
\beqn
\vert\la i\alpha\vert\la\alpha,+\vert U^\dagger_{bs}(i\frac{\pi}{4})\vert\hbox{NOON}\ra\vert^2=\frac{2^{N-1}\vert\alpha\vert^2\exp(-\vert\alpha\vert^2)}{N!\cosh\vert\alpha\vert^2}.
\eeqn
If $N$ is an  odd multiple of two, the required input state is contained as a component of the input state $\vert\alpha,-\ra\vert i\alpha\ra$.  In this case, the overlap of the ideal input with the actual input is 

\beqn
\vert\la i\alpha\vert\la\alpha,-\vert U^\dagger_{bs}(i\frac{\pi}{4})\vert\hbox{NOON}\ra\vert^2=\frac{2^{N-1}\vert\alpha\vert^2\exp(-\vert\alpha\vert^2)}{N!\sinh\vert\alpha\vert^2}.
\eeqn

\begin{figure}
\centering
\includegraphics[height=8cm,width=9cm]{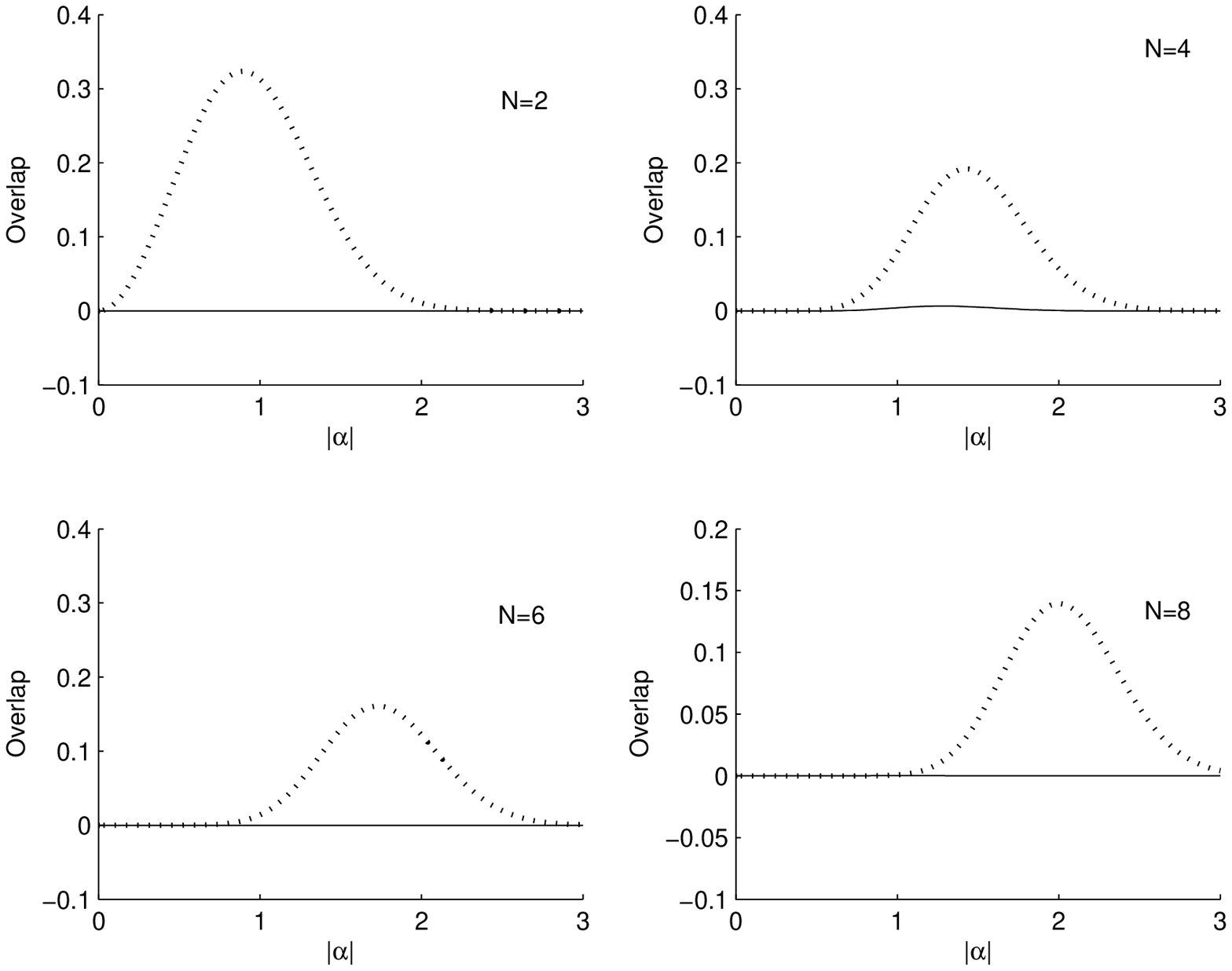}
\caption {Overlap of NOON state and the output from the beam-splitter as a function of the magnitude of the  amplitude $\alpha$, for different values of $N$.   For $N=2$ and $N=6$, the SV-CS (continuous) and the OCS-CS (dash) cases are shown.  For $N=4$ and $N=8$, the ECS-CS (dash) and the SV-CS (continuous) are shown.  The overlap in the SV-CC case is nonzero which, because of comparatively smaller magnitude than the other cases, appears to be nearly zero.  
Though the overlap is nonnegative, for the purpose of depicting near-zero values, the scale of the ordinate is extended to include negative values. }
\label{fig:FigI}
\end{figure}

In Fig. \ref{fig:FigI}  the dependence of the overlap on the amplitude $\alpha$ is shown.  The figures corresponding to $N=2$ and $N=6$ depict the overlap for the OCS-CS with the ideal input and that of the SV-CS with the ideal input.  It is seen that the overlap for the former is orders of magnitude larger than the later.     If N=4 or 8, the overlap of the ideal input state with the ECS-CS combination  is still larger than that in the SV-CS case.  In the context of experiments, these results imply that the probability of NOON state generation is correspondingly higher.   It may be pointed out that the overlap in SV-CS case is orders of magnitude smaller than the other two cases, that the corresponding curve does not appear to deviate from zero due to the scale used in the figures.
 Also, 
  the peak values of the overlap occur for progressively  larger values of $\vert\alpha\vert$ as $N$ increases.   Therefore, by properly  choosing the amplitude $\alpha$, it is possible to maximize the post-selection probability for a particular NOON state.

\section{Summary}
The NOON state fidelity of the beam-splitter output is enhanced by using the even/odd coherent states and the coherent states in the input.  
 The even coherent  states are suitable to produce the NOON states if $N$ is an even multiple of two, while the odd coherent states are the apt state if $N$ is an odd multiple of two.  The probability of NOON state detection can be optimized by proper choices for the amplitudes of the input coherent state and the even/odd coherent states.  Compared to the case of using  the squeezed vacuum in the   input, the probability of NOON state detection by post-selection is higher if superposed coherent states are used.   
The possibility of higher fidelity makes allowance for the non-ideal characteristics of the various optical components used in the experiments.  In conclusion, unit-fidelity and high-probability generation of the NOON states is better achieved by using  the even/odd coherent states.

\newpage
\end{document}